# Next-Generation MAC Technique for Priority Handling in Industrial Cyber-Physical Systems

Anwar Ahmed Khan
*Millennium Institute of Technology and Entrepreneurship, Pakistan.*
*South East Technological University, Ireland*
yrawna@yahoo.com
0000-0002-2237-5124

Farid Nait-Abdesselam
*Universite' Paris Cite',*
*France*
farid.nait-abdesselam@u-paris.fr
0000-0002-5042-5387

Indrakshi Dey
*South East Technological University*
*Waterford, Ireland*
Indrakshi.Dey@WaltonInstitute.ie
0000-0001-9669-6417

*Abstract*—Next Generation Media Access Control (NGMA) techniques have been designed to support diverse applications with heterogeneous priorities. In industrial cyber-physical systems (CPS), the number of connected devices and systems is expected to grow significantly, demanding dependable and prompt network services. In this work, we present a novel scheme, Dynamic Fragmentation-MAC (DyFrag-MAC) that offers dynamic, differentiated channel access to the traffic of various priorities. DyFrag-MAC works on fragmenting the data of normal priority in order to support early delivery of urgent priority data. In prior work, urgent priority data either had to wait for the complete transmission of lower-priority packets or relied on multi-channel protocols to gain access. We compared the proposed fragmentation scheme with FROG-MAC and industrial Deterministic and Synchronous Multi-channel Extension (i-DSME). FROG-MAC fragmented the lower priority packets, but did not adjust the fragment size dynamically, whereas i-DSME utilized multiple channels and adaptive contention mechanisms; both protocols lack the ability to preempt ongoing lower-priority transmissions. Hence, the performance evaluation in terms of average delay and throughput reveals better performance of DyFRAG-MAC for the heterogeneous traffic.

## I. Introduction

In the industrial cyber-physical systems (ICPS), ensuring efficient and reliable communication is crucial due to the increasing number of connected devices with diverse priority requirements. There are various industrial applications which generate data of multiple priorities [1]; for example, critical emergency control signals, real-time process monitoring, and supervisory control data, require low latency and high reliability, whereas the applications of predictive maintenance and routine monitoring can tolerate higher delays. Traditional MAC protocols struggle to handle heterogeneous traffic efficiently [2], often leading to high delays for urgent transmissions or inefficient resource utilization. Next Generation Multiple Access (NGMA) techniques aim to address these challenges by integrating dynamic scheduling [3], multi-channel access [4], and adaptive contention mechanisms [5]. These advancements are particularly relevant for applications that require real-time communication [6].

Multiple techniques have also been combined in MAC protocols to improve the prioritized access and differentiated Quality of Service (QoS) for industrial applications. For example, in Industrial Deterministic Synchronous Multichannel Extension (i-DSME) [7], adaptive contention mechanisms and multi-channel access are utilized to enhance network efficiency. i-DSME dynamically modifies the Contention Access Period (CAP) and contention-free period (CFP) within its super-frame to prioritize urgent traffic, while utilizing multiple channels to enhance throughput and minimize collisions.

Despite using multiple techniques of super-frame, dynamic contention and channel hopping, the MAC protocols often fail to offer a true priority to the urgent traffic, due to the ongoing transmission on the channel [8]. Most existing MAC protocols do not allow urgent traffic to preempt or interrupt the transmission of normal traffic. To overcome this challenge, we proposed fragmentation-based MAC, FROG-MAC in our previous work. FROG-MAC [9] transmits normal packets in fragments instead of a single unit; pauses are introduced between the fragments to let the urgent traffic interrupt and send its request packet. This way, the urgent packet gets a chance of quicker transmission without the need of waiting until the transmission of normal packet ends.

In this paper, we present a novel dynamic fragmentation MAC scheme for multi-priority traffic, Dynamic Fragmentation MAC (DyFrag-MAC) by enhancing our previous work FROG-MAC [9]. DyFrag-MAC dynamically modifies the fragment size of lower-priority packets to enable faster transmission of higher-priority packets, while simultaneously optimizing the use of channel resources. To the best of our knowledge, no approach like DyFrag-MAC has been proposed in the past.

The major contributions of the present work are listed below:

- To propose a dynamic fragmentation based Next Generation MAC scheme, DyFrag-MAC.
- To conduct a performance evaluation of DyFrag-MAC against FROG-MAC and i-DSME.
- To present a comparison of i-DSME, FROG-MAC and DyFrag-MAC for normal and urgent priority traffic for

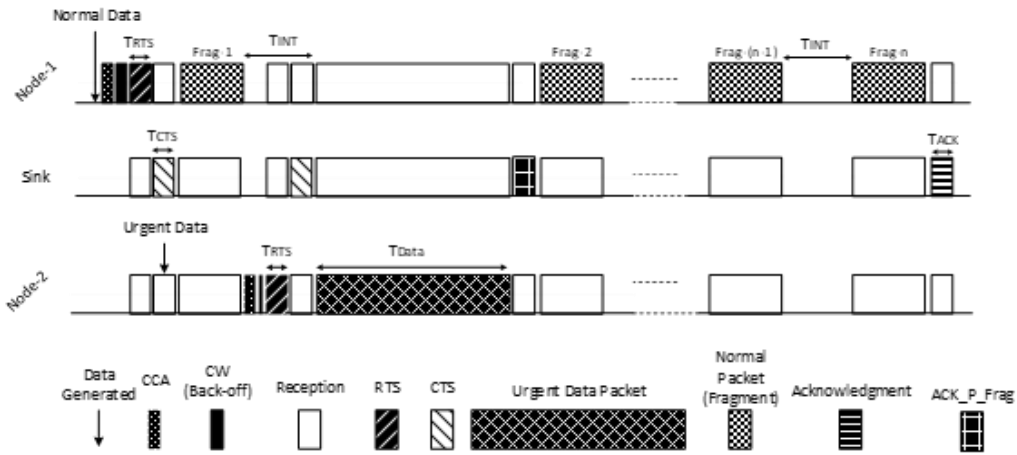

Fig. 1: Basic Operation of FROG-MAC.

illustrating the performance of each protocol for industrial CPS.

Rest of this paper has been organized as follows: Section II presents a brief overview of related work; section III details the design and operation of proposed scheme DyFrag-MAC; section IV presents the experimental setup; section V describes the results and finally, section VI concludes the paper and offers an insight into the future work directions.

## II. RELATED WORK

Adaptability has been one of the key parameters of MAC protocols, in order to serve the heterogeneous applications efficiently. Various schemes such as adaptive contention window [10], multi-priority queues [11], multi-channel usage [12] and adaptive data aggregation [13] have been used to ensure timely delivery of high priority data. Moreover, a wide range of MAC protocols embedding machine learning/deep learning functionality have also been proposed in the recent past [14]. In this section, we briefly review some of such approaches:

6TiSCH: IPv6 over the Timeslotted Channel Hopping (TSCH) mode of IEEE 802.15.4e has been a common method used for industrial automation and process control over the past decade [15]. It combines the reliability of TSCH with IPv6 networking capabilities to enable deterministic and low-power communication in CPS. 6TiSCH operates by scheduling communication slots in a time-slotted structure, ensuring minimal collisions and efficient bandwidth utilization [16]. To handle prioritized traffic, 6TiSCH allows for slot reservations and dynamic scheduling, where high-priority packets can be allocated dedicated transmission opportunities, reducing delays for time-sensitive applications. Additionally, channel hopping enhances reliability by mitigating interference; due to the determinism and reliability 6TiSCH offers, it has become a prominent choice for Industrial IoT [17].

AdaptiveHART (Adaptive Highway Addressable Remote Transducer) [18]: is an enhancement of WirelessHART [19], designed to improve real-time communication and prioritized data delivery in industrial wireless networks. WirelessHART faced challenges in handling dynamic traffic loads and ensuring low-latency transmission for critical control signals [20]. AdaptiveHART addresses these issues by dynamically adjusting time-slot allocation and transmission scheduling based on data priority. Critical messages, such as emergency shutdown signals, receive higher priority over routine monitoring data through optimized retransmission mechanisms, reduced back-off times, and dedicated time slots.

All of the above protocols rely on priority mechanisms that operate only when the channel is idle and no transmission is currently in progress [21]. However, once the nodes begin transmission, even for the lower priority data, the nodes with urgent packets have to wait as they do not have any interruption method. Although multi-channel protocols have also been developed, but they also have limitations such as channel switching delays, increased coordination overhead, and inefficient spectrum utilization in high-traffic scenarios. Additionally, many multi-channel protocols rely on pre-assigned slots or static channel allocations, which may not adapt well to dynamic traffic variations. As a result, urgent data may still experience delays, especially in scenarios where channel conditions fluctuate or contention levels are high.

In our previous work FROG-MAC [9], the approach of fragmenting lower priority/normal data was introduced, in order to offer an early transmission opportunity to the higher priority/urgent traffic. The basic operation of FROG-MAC has been illustrated in figure 1. Node-1 performs a conventional Clear Channel Assessment (CCA) and backoff before transmitting a Request to Send (RTS). Upon receiving a Clear to Send (CTS) from the sink, Node-1 begins sending its packet in fragments. However, during one of the pauses between fragments, the sink receives an RTS from another node, Node-2, indicating the presence of an urgent packet. When the sink responds with a CTS for Node-2, Node-1 also overhears this transmission and becomes aware of the urgent traffic. As a result, it temporarily pauses its fragment transmission and resumes only after the urgent packet has been successfully

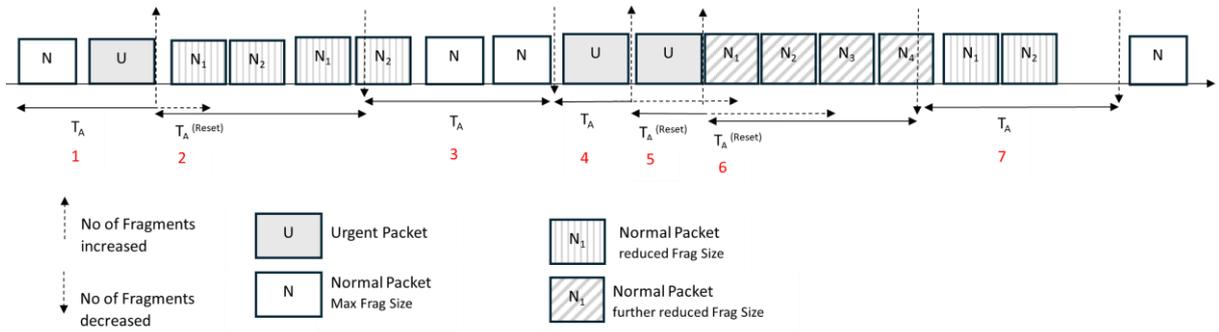

Fig. 2: Dynamic Fragmentation Operation of DyFrag-MAC.

transmitted. Further details on FROG-MAC can be found in [9].

## III. PROPOSED SCHEME: DYFRAG-MAC

In this work, we propose a Dynamic Fragmentation based MAC protocol that prioritizes high priority data by dynamically adjusting the size of low priority fragments. The frequency of incoming packets belonging to each priority is monitored, and the fragment size is adjusted accordingly; this allows the urgent traffic to interrupt the transmission quickly, reducing its waiting time. In contrast to traditional approaches that primarily optimize the performance of higher-priority data, our method also emphasizes enhancing overall channel utilization by allowing lower-priority data to use available resources when urgent traffic is absent, rather than leaving the channel idle. The basic operation of DyFrag-MAC is illustrated in figure 2.

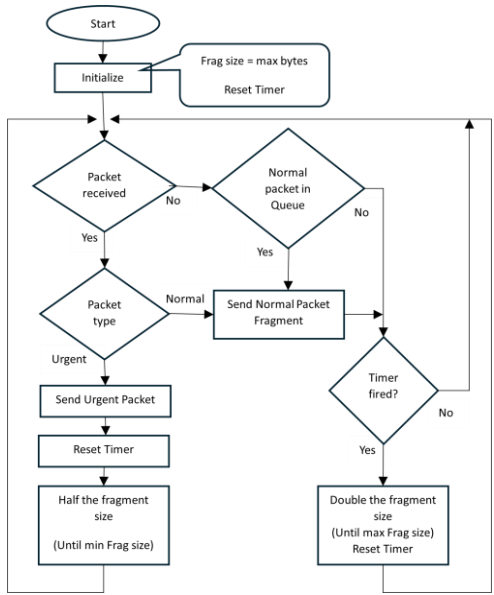

Fig. 3: Flowchart Describing the Dynamic Fragmentation Operation of DyFrag-MAC

The fragment size is dynamically adjusted by monitoring the arrival of urgent packets within defined intervals called "Assessment Cycles" ($T_A$). Initially, both normal and urgent packets remain unfragmented. However, when an urgent packet arrives, fragmentation of the normal packet begins, and the timer, $T_A$, is reset. For instance, as illustrated in figure 2, if an urgent packet arrives during the first $T_A$ cycle, the normal packet (N) is split into two fragments, $N_1$ and $N_2$. In the following cycle, since no urgent packet arrives, the number of fragments decreases until the normal packet is restored to its full size (N). During the third cycle, with no urgent packet arrivals, normal packets continue without fragmentation. However, in the fourth and fifth cycles, the arrival of two urgent packets causes the normal packet to be divided into four fragments. This process continues, increasing the number of fragments until the maximum limit is reached. Moreover, the operation of suggested dynamic fragmentation scheme has also been presented in the form of a flow chart in the figure 3.

Figure 3 shows that initially, the fragment size is kept at the maximum: no fragmentation is implemented. Once the urgent packet is received, the fragment size is halved, and the practice is continued until the minimum fragment size is reached. On the other hand, if normal packet is received, the fragment size is doubled, until the maximum fragment size is reached. Hence, the suggested approach will ensure that in case no urgent traffic is available, longer fragments are sent, reducing the delay of normal traffic.

## IV. EXPERIMENTAL SETTINGS

To evaluate the performance of the proposed DyFrag-MAC protocol, simulations were conducted using Contiki OS and its network simulator Cooja, which is widely used for testing MAC protocols in low-power and lossy networks. The simulation environment models an industrial CPS scenario, where factory equipment communicates directly with a central cluster head (sink) node for timely and reliable data reporting.

A total of 11 nodes were used in the experiments, consisting of 10 source nodes and 1 sink node. To assess scalability and performance under different network loads, the number of source nodes was varied incrementally while keeping the sink node constant. For example, when the total number of nodes was two, one node was configured as the source and the other as the sink; when there were three nodes, two served as sources and one as the sink, and so forth, up to ten source nodes and one sink node.

All simulations employed the Unit Disk Graph Medium (UDGM) with a distance loss model to represent radio propagation in a simplified manner. To ensure reproducibility and consistent comparisons across different MAC protocols, fixed random number seeds were used throughout all simulation runs. The protocols evaluated include FROG-MAC, DyFrag-MAC, and i-DSME. The key performance metrics measured were average delay and throughput.

It should be noted that this simulation environment represents an abstract and idealized model that does not fully capture real-world factors such as hardware limitations, radio interference, buffer overflows, and dynamic channel conditions. While this abstraction facilitates controlled and repeatable experiments, it may not fully reflect the behavior of the protocols under practical deployment scenarios. Addressing these limitations will be part of future work through hardware testbed evaluations and simulations incorporating more realistic models.

The next section details the simulation results and their analysis based on the experimental configuration described above.

## V. RESULTS AND DISCUSSIONS

Figure 4-a shows the variation of average delay for all the protocols, for both types of traffic with the varying number of nodes. It has been shown that as the number of nodes increases, the delay proportionally increases due to the increasing contention, packet collisions and retransmissions. The delay is found to be lowest for FROG-MAC and second lowest for DyFrag-MAC; this difference between the two protocols is observed because in the FROG-MAC, there was no dynamic adjustment of the fragment sizes which implies that the urgent traffic will be always dealt with the highest priority regardless of the possible under-utilization of the resources. However, there is not a very significant difference between the delay for urgent traffic faced by FROG-MAC and DyFrag-MAC. On the other hand, the delay of normal traffic was found to be highest for the FROG-MAC as it always sends fragments of fixed sizes, due to which the delay of FROG-MAC increases; comparatively, the delay is significantly lower for the normal traffic of DyFrag-MAC because it increases the fragment size in case no urgent traffic is available, which facilitates quicker transmission of normal traffic.

The comparison of delay for DyFrag-MAC and i-DSME also shows interesting trends. First, the delay for i-DSME is higher for the normal traffic as comapared to both FROG-MAC and DyFrag-MAC; this happens because i-DSME focuses on reducing the delay of higher priority traffic by reducing its CAP, and allowing it to access multiple channels during CFP. However, it increases the delay for normal traffic, which could not transmit even if the slots are available. Moreover, the delay for urgent traffic in i-DSME is also higher due to limitations in its prioritization mechanisms. Despite leveraging adaptive contention and multi-channel access, i-DSME lacks the ability to preempt or interrupt ongoing lower-priority transmissions. Consequently, once the transmission of normal traffic begins, higher-priority packets must wait for the ongoing transmission to complete, leading to increased delays. Additionally, the reliance on predefined slot allocation and limited flexibility in slot reassignment further restricts its ability to dynamically accommodate urgent traffic, making it less efficient in handling real-time or latency-sensitive applications.

We also evaluated the impact of varying the fragment size. The comparison of FROG-MAC in figures 4-a and 4-b shows that when the fragment size is reduced to 2, the delay for normal traffic increases significantly, whereas it reduces for the urgent traffic. This occurs because reducing the fragment size increases the transmission time for normal traffic, allowing urgent packets to interrupt the channel more rapidly. On the other hand, there is no impact on the results of DyFrag-MAC and i-DSME, as fragment size was not varied in these protocols.

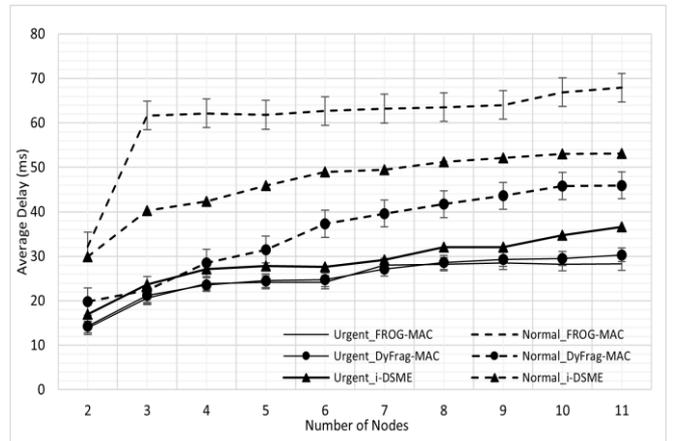

(a)

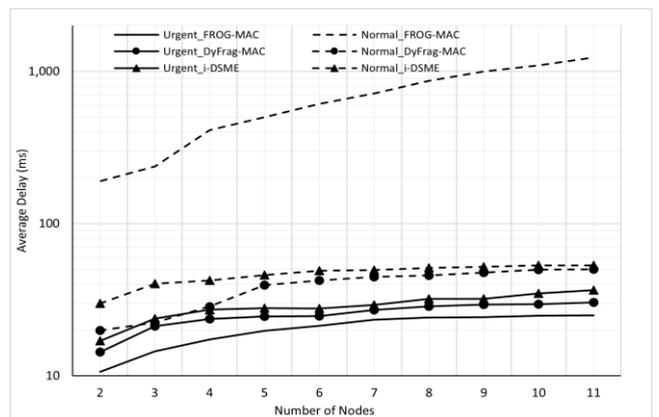

(b)

Fig. 4: Delay Comparison. (a) Fragment Size= 16, (b) Fragment Size= 2

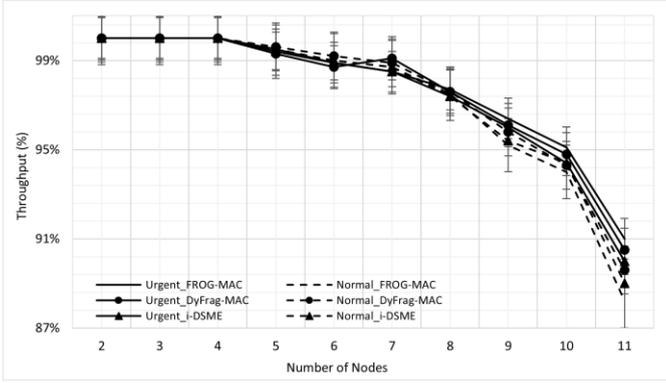

(a)

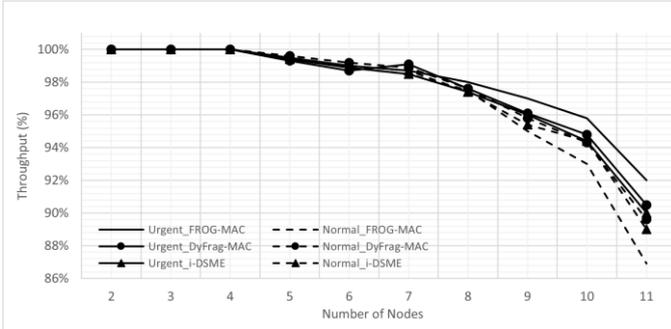

(b)

Fig. 5: Throughput Comparison. (a) Fragment Size= 16, (b) Fragment Size= 2

Figure 5 shows the impact of varying number of nodes on the throughput, where figure 5-a presents the case for fragment size 16, and 5-b shows fragment size 2. As the number of nodes increases, the throughput reduces for all the protocols. Highest throughput has been achieved for FROG-MAC with urgent traffic, whereas the lowest has been observed for normal traffic with FROG-MAC. Furthermore, DyFrag-MAC achieves higher throughput for both normal and urgent traffic compared to i-DSME, thanks to more efficient resource utilization enabled by dynamic fragmentation.

## VI. Conclusion and Future Work

In this paper, we presented a dynamic fragmentation based MAC protocol, DyFrag-MAC. The protocol, due to its dynamic adaptation to the incoming traffic scenarios, fits well to the requirements of industrial CPS. DyFrag-MAC fragments the normal packets to offer pauses in between to the urgent traffic; the size of fragment is dynamically adjusted based on the incoming arrival frequencies. The dynamic adjustment facilitates efficient usage of channel resources while not compromising the delay of urgent traffic. The performance of DyFrag-MAC has been compared with FROG-MAC and i-DSME and DyFrag-MAC provided better results in terms of delay and throughput.

In future, this work can be extended in various directions: DyFrag-MAC can be further improved by incorporating machine learning-based traffic prediction models to enhance the dynamic adaptation of fragmentation sizes based on real-time traffic patterns. Additionally, its performance can be evaluated in multi-hop industrial networks to analyze the impact on end-to-end latency and reliability. Another important extension is the integration of DyFrag-MAC with emerging wireless technologies such as time-sensitive networking to support ultra-reliable low-latency communication in industrial CPS. Furthermore, real-world validation through hardware implementations and testbed experiments can provide deeper insights into energy consumption, synchronization challenges, and practical deployment considerations.

## VII. Acknowledgment

This contribution is supported by HORIZON-MSCA-2022-SE-01-01 project COALESCE under Grant Number 10113073.